\documentclass[aps,prb,twocolumn,groupedaddress,showpacs,showkeys]{revtex4}

\pdfoutput=1
\usepackage{xspace}
\usepackage{graphicx}
\usepackage{epsfig}
\usepackage{amsfonts}

\begin{document}


\title{\textsf{Zener Tunneling in Semiconducting Nanotube and Graphene Nanoribbon p-n Junctions}}

\author{Debdeep Jena, Tian Fang, Qin Zhang, \& Huili Xing}
\affiliation{Department of Electrical Engineering \\
            University of Notre Dame \\
            IN, 46556}
\date{\today}

\begin{abstract}
A theory is developed for interband tunneling in semiconducting carbon nanotube and graphene nanoribbon p-n junction diodes.  Characteristic length and energy scales that dictate the tunneling probabilities and currents are evaluated.  By comparing the Zener tunneling processes in these structures to traditional group IV and III-V semiconductors, it is proved that for identical bandgaps, carbon based 1D structures have higher tunneling probabilities.  The high tunneling current magnitudes for 1D carbon structures suggest the distinct feasibility of high-performance tunneling-based field-effect transistors.
\end{abstract}

\pacs{73.40.-c, 73.40.Kp, 73.40.Lq} 
\keywords{interband, Zener, tunneling, graphene, nanoribbon, carbon, nanotube, diode, transistor}

\maketitle


Carbon-based 1D materials such as nanotubes (CNTs) and graphene nanoribbons (GNRs) are currently under extensive investigation for the novel fundamental physics they exhibit, as well as possible applications they might have in the future\cite{avouris, geim}.  A large class of traditional semiconductor devices rely on the quantum mechanical tunneling of carries through classically forbidden barriers.  Among these, the Esaki diode, resonant tunneling diode, and backward diode are the prime examples\cite{szeTunneling}.  In addition, the high-field electrical breakdown in a number of semiconductors occurs by interband Zener tunneling.  The phenomena of tunneling has been studied extensively for traditional parabolic-bandgap semiconductors in 3D bulk, as well as quasi-2D and quasi-1D heterostructures.  Semiconducting CNTs and GNRs do not have parabolic bandstructures, and the carrier transport in them approaches the ideal 1D case.  In that light, it is timely to examine the phenomena of tunneling in these materials.  

Tunneling currents in semiconducting CNT p-n junctions have been measured and analyzed recently (see \cite{appenzeller1, appenzeller2, mceuen}).  For tunneling probabilities, an energy-dependent carrier effective mass has been used earlier\cite{bachtold}, to take advantage of previously existing results of parabolic bandstructure semiconductors.  In this work, we evaluate the Zener tunneling probabilities of CNT and GNR based p-n diodes starting from their intrinsic bandstructures, which removes the need to define an effective mass.  In addition to interband tunneling probabilities, a number of fundamental associated parameters characterizing the tunneling process are found.

The bandstructure of the nth subband of a semiconducting CNT or GNR is given by\cite{tian}
\begin{equation}
\mathcal{E} = s \hbar v_{F} \sqrt{ k_{x}^{2}  + k_{n}^{2}},
\label{bandstructure}
\end{equation}
where $2 \pi \hbar$ is the Planck's constant, and $v_{F} \sim 10^{8}$ cm/s is the Fermi velocity characterizing the bandstructure of graphene.  $s= + 1$ denotes the conduction band, and $s = -1$ denotes the valence band.  The electron momentum along the CNT or GNR axis is $\hbar k_{x}$.   

For GNRs, the transverse momentum is quantized by the ribbon width\cite{breyfertig}; $k_{n} = n \pi / 3 W$ where $n = \pm 1, \pm 2, \pm 4, \pm 5, \pm 7, \pm 8...$ for a GNR of dimensions $(x, y) = (L, W)$ where $W << L$.  The corresponding bandgap  is $\mathcal{E}_{g} = 2 \hbar v_{F} k_{1} = 2 \pi \hbar v_{F} / 3 W \sim 1.3/W$ eV (where $W$ is in nm).  For comparison, a semiconducting CNT of diameter $D$ has the same bandstructure with $k_{1} = 2/3D$, and a bandgap of $\mathcal{E}_{g} = 4 \hbar v_{F} / 3 D \sim 0.8/D$ eV, where $D$ is in nm.  If $W = \pi D / 2$, the properties (bandgap, bandstructure) of semiconducting CNTs and GNRs are similar. The results derived below are applicable to GNRs and CNTs on equal footing.  It is assumed that the length of the GNR (CNT) is much larger than the width (diameter) such that the longitudinal momentum of carriers in the ribbon are quasi-continuous.

\begin{figure}
\begin{center}
\leavevmode \epsfxsize=1.7in 
\epsfbox{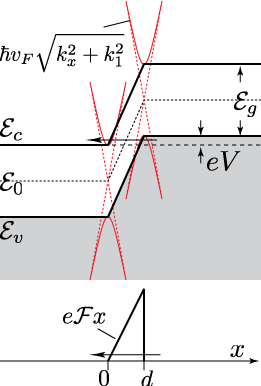} 
\caption{Interband tunneling in a reverse-biased GNR/CNT p-n junction, and the potential barrier seen by tunneling electrons.}
\label{btbt}
\end{center}
\end{figure}

We now evaluate the interband tunneling probability in a $n^{+} - p^{+}$ GNR or CNT diode of bandgap $\mathcal{E}_{g}$.  We consider that the doping of the n and p-sides are such that the equilibrium Fermi level is at the conduction band edge ($\mathcal{E}_{c}$) in the n-side, and at the valence band edge ($\mathcal{E}_{v}$) on the p-side.  Such doping could be either chemical, or electrostatic\cite{avouris}.  Under this situation, a forward bias would not lead to current flow (this is similar to the `backward diode'\cite{szeBackwardDiode}).  When a reverse bias $e V$ is applied, the band diagram looks as shown in Figure \ref{btbt}.  Let the electric field in the junction region be $\mathcal{F}$.  We assume that the depletion region thickness, and the net electric field does not change appreciably from the equilibrium values (true under small bias voltages).  Then the potential energy barrier seen by electrons in the valence band of the $p^{+}$ side is
\begin{equation}
V_{0} (x)  =  e \mathcal{F} x,
\end{equation}
in the range $ 0 < x < d$, such that $ d = \mathcal{E}_{g} / e \mathcal{F}$.  This is indicated in Figure \ref{btbt}.  $\mathcal{E}_{0}$ is the Dirac point of the underlying graphene bandstructure, for both GNRs and CNTs, and serves as a convenient reference of energy in the problem.

Since the energy of carriers near the band edge ($k_{x} \approx 0$) is conserved during the tunneling of electrons from the valence to the conduction band, the condition
\begin{equation}
- \sqrt{ (\hbar v_{F} k_{x})^{2}  + (\frac{\mathcal{E}_{g}}{2})^{2} } + \mathcal{E}_{g} - e \mathcal{F} x = +  \sqrt{ (\hbar v_{F} k_{x})^{2}  + (\frac{\mathcal{E}_{g}}{2})^{2} }
\label{enerconv}
\end{equation}
holds for the wavevector $k_{x}$ at all $x$.  Within the tunneling barrier, the wavevector is imaginary.  Denoting this by $k_{x} = i \kappa_{x}$ where $\kappa_{x}$ is real, we obtain
\begin{equation}
\kappa_{x} (x) = k_{0} \sqrt{ 1 - (1 - \frac{x}{d})^{2} },
\end{equation}
where $k_{0} = \mathcal{E}_{g} / 2 \hbar v_{F} $ is a characteristic wavevector for tunneling.  Since $ \int_{0}^{d}   \sqrt{ 1 - (1 - \frac{x}{d})^{2} } =  \pi d / 4 $, the WKB band-to-band tunneling (BTBT) probability for the $p^{+} - n^{+}$ junction given by $T \sim \exp{ [- 2 | \int_{0}^{+d} \kappa_{x}(x) dx | ] }$ leads to 
\begin{equation}
T_{WKB} (k_{x} \approx 0) \sim \exp \big[ { - \frac{\pi}{4} \cdot  \frac{\mathcal{E}_{g}^{2}}{ \hbar v_{F} e \mathcal{F}} } \big],
\label{probNoKE}
\end{equation}

which can be expressed as $T_{WKB}  \sim \exp(- \mathcal{F}_{0} / \mathcal{F} )$, where $\mathcal{F}_{0} = \pi \mathcal{E}_{g}^{2} / 4 e \hbar v_{F}$ denotes the characteristic electric field at the junction for the onset of strong tunneling.   The corresponding characteristic barrier thickness for the onset of strong tunneling is $d_{0} \sim \mathcal{E}_{g} / e \mathcal{F}_{0} =  4 \hbar v_{F} / \pi \mathcal{E}_{g}$.

Using the value of the Fermi velocity, the characteristic field evaluates to $\mathcal{F}_{0} \sim 12.6 \times (\mathcal{E}_{g})^{2} $ MV/cm, and the characteristic tunneling distance evaluates to $d_{0} \sim 0.8/\mathcal{E}_{g}$ nm, where $\mathcal{E}_{g}$ is the bandgap of the GNR or CNT expressed in eV in both these expressions.  As an example, for a GNR with $W=5$ nm, $\mathcal{E}_{g} \sim 0.275$ eV, the characteristic tunneling field is $\sim$ 0.9 MV/cm, and the characteristic tunneling barrier thickness is $\sim$ 11 nm.

The tunneling probabilities evaluated above are for the first subband ($n=1$).  For the nth conduction and valence subbands of CNTs and GNRs, the effective subband gap scales as $\sim n \mathcal{E}_{g}$.  The tunneling probabilities of carriers near the respective subband-edges is then given by $T_{WKB, n} \sim \exp (- \pi \hbar v_{F} k_{n}^{2} / \mathcal{F} )$, which decay as $\exp(-n^{2} )$, indicating a rather strong damping the tunneling probabilities of higher subbands.  This result turns out to be identical to one for the nth transverse mode of a zero-gap 2D graphene p-n junction, as first evaluated by Cheianov and Fal'ko (see \cite{falko}).  Except for the narrowest bandgap CNTs and GNRs, the tunneling is primarily from the 1st valence subband in the p-side to the 1st conduction subband on the n-side.

Note that the above derivation uses a triangular barrier approximation.  It has been shown by Kane\cite{kane} that a parabolic barrier more accurately represents the physics of the tunneling process, and leads to an exponential factor with a different coefficient than from the triangular barrier approximation.  The difference is small, for the rest of this work, we will use the result derived above.

\begin{figure}
\begin{center}
\leavevmode \epsfxsize=3.2in 
\epsfbox{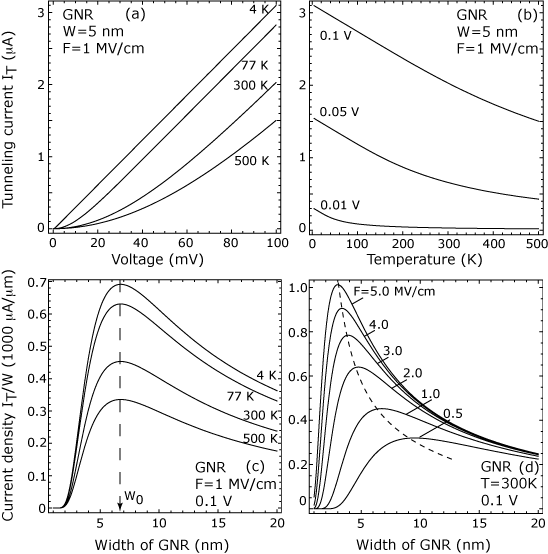} 
\caption{Tunneling currents in GNR p-n junctions: (a) $W=$5 nm device, voltage dependence at various temperatures, (b) temperature dependence at various voltages.  Tunneling currents per unit width for different GNR widths at (c) various temperatures at $\mathcal{F} = 1$ MV/cm, and (d) for various $\mathcal{F}$ at 300 K.  To maximize the tunneling current density, an optimum ($\mathcal{F}, W$) combination exists, as illustrated by the dashed lines.}
\label{currents}
\end{center}
\end{figure}

The tunneling current may now be written in the form $I_{T} = e (g_{s} g_{v} / L) \sum_{k_{x}} [ f_{v} - f_{c} ] T_{WKB} \times v_{g} (k_{x}) $, where $g_{s} = 2$ is the spin degeneracy, $g_{v}$ is the valley degeneracy (= 2 for CNTs and = 1 for GNRs\cite{avouris}), $L$ is the length of the CNT or GNR, $f_{v} , f_{c} $ are the occupation functions of the valence band states in the p-type side and the conduction band in the n-side respectively, and $v_{g} (k_{x})$ is the group velocity.  This sum may be converted into the equivalent integral
\begin{equation}
I_{T} = \frac{2 g_{v} e }{h} \int_{0}^{e V} [ f_{v}( \mathcal{E} ) - f_{c}( \mathcal{E} ) ] T_{WKB} d\mathcal{E},
\end{equation}
where $f_{v}(\mathcal{E}) = 1/(1 + \exp{[ ( \mathcal{E} - e V )/ k_{B}T ] })$ and $f_{c} ( \mathcal{E} ) = 1/( 1 + \exp{ [ \mathcal{E} / k_{B}T ] } )  $.  Here $k_{B}$ is the Boltzmann constant.  For the band diagram shown in Figure \ref{btbt}, the net tunneling current evaluates to
\begin{equation}
I_{T} = \frac{2 g_{v} e^{2}}{h} T_{WKB} \times V_{T} \ln{ [ \frac{1}{2} ( 1 + \cosh{ \frac{V}{V_{T}} } )  ]  },
\end{equation}
where $V_{T} = k_{B} T / e$.  This expression captures the temperature and bias voltage dependence of the tunneling current.  If the applied bias is much greater than the thermal energy ($V >> V_{T}$), the current reduces to a form similar to the Landauer expression $I_{T} \approx (2 g_{v} e^{2} / h) T_{WKB} ( V - V_{T}\ln{4})$.  Thus, the tunneling current has a negative temperature coefficient at high bias conditions.  The dependence of tunneling currents in GNRs on voltage and temperature are illustrated in Figure \ref{currents} (a, b).  

The tunneling current {\em per unit width} of GNRs is maximized for $W_{0} = \sqrt{ 2 \pi^{3} \hbar v_{F} / 9 e \mathcal{F} }$. For example, for $\mathcal{F} \sim 1$ MV/cm, the GNR width for maximum current drive is $W_{0} \sim 6.5 $ nm, and the current density at that width is $\sim$ 450 $\mu$A/$\mu$m, comparable to traditional FETs.  For thinner ribbons at higher fields, the current densities can approach 1000 $\mu$A/$\mu$m.  The dependence of the current densities on temperature, electric field, and GNR widths are shown in Figure \ref{currents} (c, d).

The BTBT probability in traditional parabolic-bandstructure semiconductors in the triangular barrier approximation depends on the bandgap as\cite{szeTunneling}  $T_{parabolic} \sim \exp{ ( -  4 \sqrt{ 2 m^{\star}} \mathcal{E}_{g}^{3/2}  / 3 e \hbar \mathcal{F}  ) }$, where $m^{\star}$ is a reduced carrier effective mass.  The BTBT probability for GNRs and CNTs retains the same dependence on electric field, but due to the difference in bandstructure there is a stronger dependence on the bandgap.  The effective mass does appear in the tunneling probability of the GNR or CNT diodes since their bandstructure is not parabolic at any energy.  If one compares the Zener tunneling probabilities in diodes made of CNTs or GNRs with other direct-bandgap semiconductors of the {\em same} bandgap, then the ratio 
\begin{equation}
\frac{ T_{carbon} }{ T_{parabolic} } \sim \exp{  \big[ -  \frac{ \mathcal{E}_{g}^{3/2} }{ e \hbar \mathcal{F} } ( \frac{ \pi \sqrt{ \mathcal{E}_{g} } }{ 4 v_{F} } - \frac{ 4 \sqrt{2 m^{\star} } }{ 3 }  ) \big]  }  
\end{equation}
indicates that the GNR or CNT p-n diode will have a higher interband tunneling probability if the relation $ \mathcal{E}_{g} < (16 v_{F} / 3 \pi)^{2} \times 2 m^{\star} $ is satisfied.  From the ${\bf k \cdot p}$ theory for traditional direct-bandgap semiconductors, the electron effective mass (in the conduction band) is related to the bandgap by the approximation\cite{cardona} $m^{\star}_{c} \approx (\mathcal{E}_{g}/20) m_{0}$, where the bandgap is in eV and $m_{0}$ is the free electron mass.  This leads to the requirement 
\begin{eqnarray}
m_{0} v_{F}^{2} > (\frac{3 \pi }{ 16 })^{2} \times 10 \mbox{ eV},
\end{eqnarray}
which is {\em always} satisfied since the LHS is $m_{0} v_{F}^{2} \approx 5.7 $ eV and the RHS is 3.5 eV.  Thus, the CNT or GNR p-n diode will {\em always} have a higher reverse-bias Zener tunneling probability than a traditional semiconductor of the same bandgap.  

Two more facts tilt the tunneling probability decisively in favor of CNTs and GNRs.  First, for bulk 3D p-n junctions, the transverse kinetic energy of carriers can be large, and leads to a further exponential decrease of carrier tunneling probability \cite{szeTunneling}, which is avoided in 1D structures.  Second, if normal parabolic bandgap semiconductors are shrunk to length scales comparable to that of CNTs and GNRs, their bandgap and the effective masses increase further due to quantum confinement.  

Though Zener tunneling currents are detrimental in traditional devices such as rectifiers, field-effect and bipolar transistors, it is important to note that the fundamental switching action in these devices is controlled by thermionic emission over barriers, which require a minimum of $(k_{B}T/e) \ln(10) \sim 60$ mV per decade change of current at 300 K (the `classical' limit).  However, a new crop of tunneling-FETs have been recently proposed and demonstrated\cite{appenzeller1, qin, tsujaeking}), which rely on the very mechanism studied in this work.  These devices are capable of reaching far below the classical limit for switching by exploiting quantum mechanical tunneling.  It is for such devices that high interband tunneling current drives in carbon-based 1D nanostructures hold a distinct advantage, and much promise in the future.




We gratefully acknowledge financial support from NSF awards DMR-06545698 \& ECCS-0802125 \& from the Nanoelectronics Research Initiative (NRI) through the Midwest Institute for Nanoelectronics Discovery (MIND), and acknowledge discussions with Joerg Appenzeller \& Eric Pop for this work.


\end{document}